\documentclass[11pt,a4paper,twocolumn]{article}
\usepackage[top=1.0cm, bottom=1.5cm, left=1.5cm, right=1.5cm, a4paper]{geometry}
\usepackage{authblk}
\usepackage[utf8]{inputenc}
\usepackage[english]{babel}
\usepackage[leftcaption]{sidecap}
\usepackage{url}
\usepackage{graphicx}
\usepackage{cite}

\usepackage{xcolor}
\newcommand{\yl}[1]{\textcolor{black}{#1}}

\makeatletter
\renewcommand\AB@affilsepx{, \protect\Affilfont}
\makeatother

\let\OLDthebibliography\thebibliography
\renewcommand\thebibliography[1]{
  \OLDthebibliography{#1}
  \setlength{\parskip}{0pt}
  \setlength{\itemsep}{0pt plus 0.3ex}
}

\title{
Schottky detection techniques for ultra-rare short-lived ions in heavy ion storage rings
}

\author[1]{M.~S.~Sanjari}
\author[1, 2]{Yu.~A.~Litvinov}
\author[3, 4]{G.~W.~Hudson-Chang}
\author[5]{S.~Naimi}
\author[6]{D.~Dmytriiev}
\author[1]{J.~Glorius}
\author[7]{E.~Menz}
\author[4]{T.~Ohnishi}
\author[3]{Zs.~Podolyak}
\author[1, 8]{Th.~Stöhlker}
\author[9]{T.~Yamaguchi}
\author[4]{Y.~Yamaguchi}
\author[4, 10]{A.~Yano} 
 \affil[1]{GSI, Darmstadt, Germany}
\affil[2]{HFHF, GSI, Darmstadt, Germany}
\affil[3]{University of Surrey, Surrey, UK}
\affil[4]{RIKEN, Wako, Saitama, Japan}
\affil[5]{IJCLab, Orsay, France}
\affil[6]{DESY, Zeuthen, Germany}
\affil[7]{University of Cologne, Cologne, Germany}
\affil[8]{Helmholtz-Institut Jena, Jena, Germany}
\affil[9]{Saitama University, Saitama, Japan}
\affil[10]{University of Tsukuba, Ibaraki, Japan}

\date{}

\begin{document}

\maketitle

\begin{abstract}
  Non-destructive Schottky detectors are indispensable devices widely used in experiments at heavy-ion storage rings. In particular, they can be used to accurately determine the masses and lifetimes of short-lived exotic nuclear species. Single-ion sensitivity -- which enables highest sensitivity -- has been regularly achieved in the past utilizing resonant cavity detectors. Recent designs and analysis methods aim at pushing the limits of measurement accuracy by increasing the dimensionality of the acquired data, namely the position of the particle as well as the phase difference between several detectors. This work describes current methods and future perspectives of Schottky detection techniques focusing at their application for mass and lifetime measurements of the most rare and simultaneously short-lived radio nuclides.
\end{abstract}

\section{Schottky detection technique}

\yl {Swift charged} particles traversing accelerator chambers induce an opposite equivalent charge on the inner surface of the beam pipe. For rapidly moving particles, this charge is assumed to be concentrated around an infinitesimally thin ring along the particle's trajectory. On the surface of an isolated region within the beam pipe, such as detector plates (see e.g., \cite{schaaf1991}), the surface charge undergoes redistribution until equilibrium is achieved. For a particle passing repeatedly, this charge redistribution can be quantified as an equivalent induced current $i(t)$, characterised by the shape of a periodic delta function for a single particle \cite{chattopadhyay1984}.

The same phenomenon occurs on the inner walls of a cavity; however, the duration of the oscillation of the charge redistribution exceeds the time required for the particle's passage. This extended duration arises because the components of the cavity walls facing each other form a transient oscillating electric dipole. This dipole generates an alternating magnetic field, resulting in the continuous exchange of energy between the stored electric and magnetic fields. The measurement is typically conducted by extracting the field energy using either an electric pin or a magnetic loop, the shape and position of which has a significant impact on the efficiency of extracted signal.

With many particles circulating around the storage ring, the current $i(t)$ will show a macroscopic DC value known as \textit{beam current} $I_B$. The spectrum of this current repeats its shape at integer multiples (harmonics, denoted as $h$) of the particle's revolution frequency, commonly referred to as \textit{Schottky bands}. The total power, which is equal for all bands around the harmonics will be \cite{caspers2008}
\begin{equation}
\langle I\rangle^2=2(qe)^2f_r^2N
\label{eqn:schottkycurrent}
\end{equation}
where \yl{$q$ and $f_r$ are the charge state and the revolution frequency of the ions, respectively, $N$ their number,} and $e$ the elementary charge.

\yl{Particles with slightly different magnetic rigidities, $Br=mv\gamma/q$, where $m$, $v$ and $\gamma$ stand for the particle mass, velocity and the relativistic Lorentz factor, respectively, show up in the revolution frequency spectra as closely lying peaks}. With increasing harmonic number, \yl{h}, one can observe that the distance between these peaks, \yl{$\Delta{f}$}, \yl{increases as $h\cdot\Delta{f}$}. This is known as Schottky \textit{band broadening} which comes at the cost of losing peak power, but is in fact a desirable feature when it comes to \yl{{\it fast}} resolving nearby peaks. This is often the case for the spectrum of \yl{short-lived low-lying nuclear} isomeric states. Cavity based Schottky detectors exploit this feature by enhancing the signal at higher frequencies. Successful examples of such cavity based Schottky detectors can be found in \cite{nolden2011,sanjari2013,wu2013,suzaki2015,wang2025,yamaguchi2016,sanjari2020}
.

\subsection{\yl{Storage ring mass and lifetime spectrometry}}

Time resolved frequency analysis can be used for the measurement of \yl{masses and lifetimes of} unstable \yl{nuclear species}. Mass measurement is done by comparing the revolution frequencies of \yl{nuclei with} unknown \yl{masses to the ones with known masses}, according to the governing equation \cite{fgm, litvinov2013}:

\begin{equation}
\frac{\Delta f}{f}=-\frac{1}{\gamma_t^2}\frac{\Delta (m/q)}{(m/q)}+\left(1-\frac{\gamma^2}{\gamma_t^2}\right) \frac{\Delta v}{v}
\label{eqn:mass}
\end{equation}
where $\gamma_t$ is known as \textit{transition} \yl{point} which \yl{is} related to storage ring's momentum compaction factor $\alpha_p$.

In order to increase the measurement accuracy, the velocity spread of the particles ($\Delta v$ in equation \ref{eqn:mass}) must be addressed. \yl{One of the ways is to reduce the velocity spread to a comparatively small value by applying beam cooling methods \cite{steck2020}}. However, since the cooling \yl{takes some time to reach the required beam quality}, efforts have been made to perform mass and lifetime measurements of shorter-lived states by tuning the lattice of the storage ring to the isochronous ion-optical (ISO) mode, where $\gamma = \gamma_t$ for the desired species \yl{\cite{hausmann2000, hausmann2001}}. In the ISO mode, the magnetic lattice of the ring is tuned such that the velocity spread of particles of different momenta is compensated by the difference in their orbits lengths, so that the resulting flight times or revolution frequencies are equal \yl{\cite{wollnik2015}}. This \yl{so-termed} Isochronous Mass Spectroscopy (IMS) method is often performed using time of flight detectors (TOF) \yl{\cite{zhang2016,yamaguchi2021}.} Recently a combined Schottky and isochronous mass (and lifetime) spectroscopy (S+IMS) method has been developed and successfully implemented where the advantages of isochronous mode was brought together with the benefits of non-destructive detection using resonant Schottky cavity detectors \yl{\cite{tu2018,freire2024}}.

For lifetime measurements, equation \ref{eqn:schottkycurrent} together with time resolved frequency analysis, allows for monitoring the decrease of spectral power (i.e. the area under the power peak), and with that, lifetime of unstable particles \yl{\cite{litvinov2011, bosch2013,leckenby2024,sidhu2024}}. For individual particles that leave distinct decay patterns, the decay times can be determined individually (for example in \cite{kienle2013}). For few particles, an spectral add up method can be used \cite{freire2024,sanjari2025}.

\subsection{Towards higher accuracies}

Further increase in measurement accuracy puts stronger demands on reduction of uncertainties of the \yl{term containing the} velocity spread in equation \ref{eqn:mass}. The cooling method, \yl{although very powerful for the longer-lived species, is too slow for application to short-lived nuclei of interest today}. Neither does the isochronous setting allow for a complete elimination of the velocity spread. \yl{Here, the advantage of the storage ring to be capable to simultaneously cover a broad range of various species brings in a complication. The bandwidth of the measured $m/q$ ratios is several percents. Therefore, to match the given set magnetic rigidity, the ion velocities have to vary by several percents as well. Hence,} apart from the extreme vicinity around the particle species whose \yl{mean $\gamma$} has been chosen to tune the lattice to the isochronous mode, all other particles show peak broadening due to the effect of \yl{the mismatch of their $\gamma\neq\gamma_t$ \cite{geissel2006}}. 
\yl{This behaviour has been onserved in all storage rings employing isochronous mode for mass measurements, namely R3 in Japan \cite{abe2024} and CSRe in China \cite{wang2023,yu2024}.} This \textit{anisochronousity} effect, as it is sometimes referred to, needs to be dealt with, when aiming for higher accuracies.

The alternative to the elimination of the velocity \yl{mismatch} would be the accurate determination of its value. In order to achieve this goal, the dimensionality of the data needs to be increased \yl{\cite{geissel2005}}. The existing three dimensions of power, frequency and time, could be augmented by considering the phase difference and position of the particles, each of which would allow for the direct or indirect determination of the velocity of the particles.

The phase difference considers the signal correlation between two or several detectors and is currently being investigated \cite{wang2025}. In the following we expand on how the position \yl{determination} can be used for the reduction of uncertainties.

\subsection{Position sensitive Schottky detectors}

Position sensitive Schottky detectors of cavity based and non-cavity based types have been utilised throughout the history of the particle accelerators \cite{chen2014hist}. Among the cavity based detectors, those which utilise the dipole mode show small sensitivity around the center of beam pipe of storage rings with large beam pipe apertures.

A novel design with an elliptical shape was proposed in order to circumvent this problem by the use of the strong monopole mode. Particles passing through the offset beam pipe strongly couple to the monopole mode, albeit with different intensities depending on their offset \cite{sanjari2014,sanjari2015}. Normalised to the signal of a reference cavity with otherwise identical characteristics, the combined signal from a Schottky Cavity Doublet (SCD) can be used in order to extract the position information. This new dimension of data can be utilised in order to correct for the magnetic measured magnetic rigidity of the particles.

\begin{figure}
\includegraphics[width=0.5\textwidth]{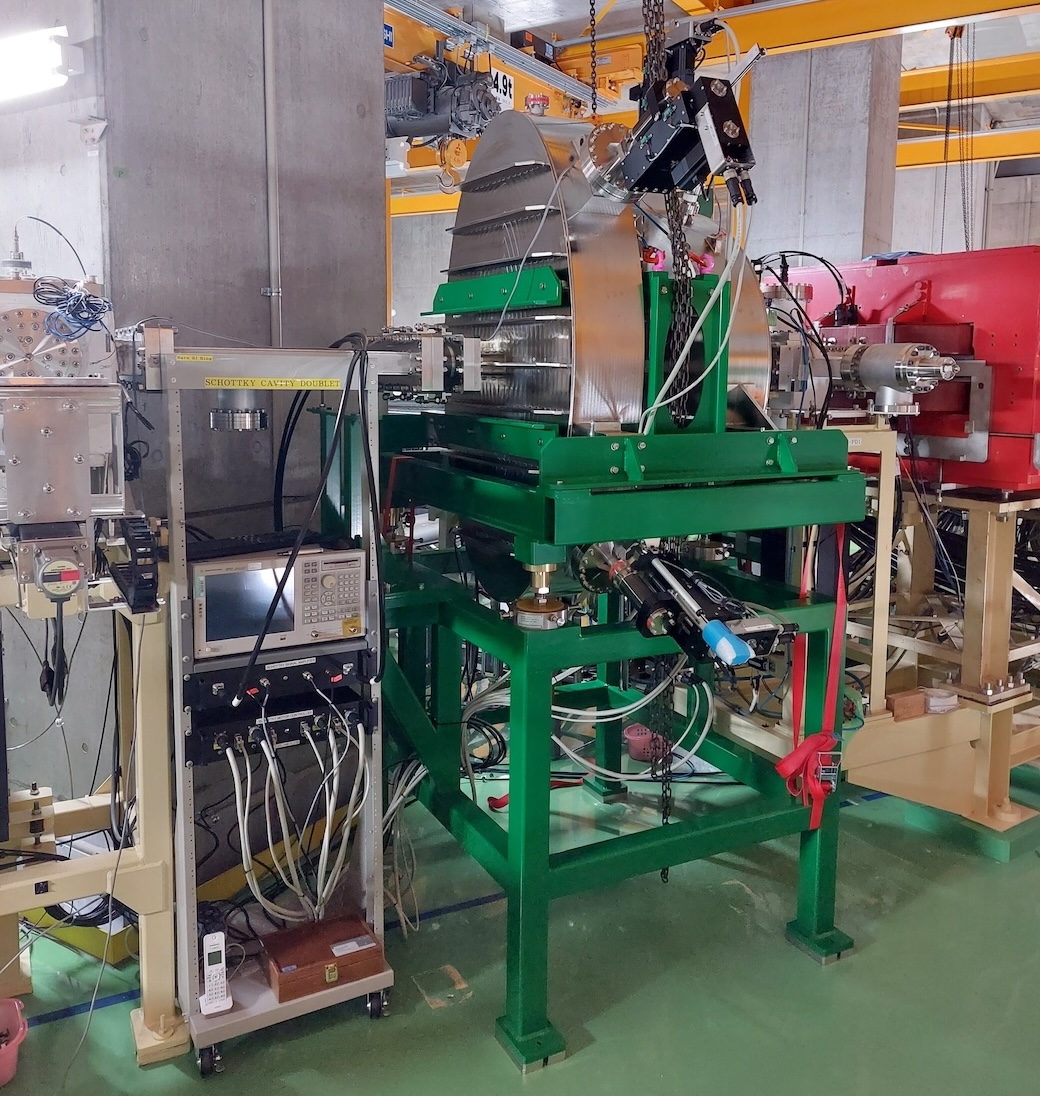}
\caption{
\yl{A photograph of a Schottky Cavity Doublet (SCD) installed at R3 storage ring. It is} shown above on top of green mechanical support (elliptical cavity left). A separate publication will be dedicated to the technical details and features of the new detector. (Photo: S. Sanjari)}
\label{fig:r3scd}
\end{figure}

In order to test this principle, an SCD was designed for the R3 storage ring \cite{dmytriiev2020}. Mechanical construction and manufacturing was carried out at GSI Darmstadt. The installation in R3 storage ring has been accomplished (see figure \ref{fig:r3scd}).

\subsection{Correction method}

The proposed correction method for mass determination is similar to the existing $B\rho$-IMS method described in \cite{wang2022,zhang2023}, but unlike the case of TOF detectors it takes advantage of the non-destructive features of the SCD. This allows for simultaneus measurement of masses and lifetimes \cite{litvinov2023}.

In the isochronous mode, each particle with the same magnetic rigidity will have the same path length, \yl{$C$}, and with that the same lateral offset, i.e. $\rho$ at the location of SCD, which is deliberately installed at a highly dispersive section. Simulations confirm, that particles with slightly different momenta will occupy different orbits \cite{hudsonchang2025}. Using the SCD, a position value is assigned to every measured ionic peak in the revolution frequency spectrum \cite{chen2015,chen2016}. The magnetic \yl{rigidity} is determined as function of path length \yl{$B\rho(C)$} for the reference particles only. A fit over the resulting function can then be used to determine the $m/q$ of the unknown particles. \yl{Since the measurements in the ring are repeated many times by injecting new freshly produced particles, the spectra will contain random nuclear species. In order to explore all possible correlations in these spectra, all information will be put into a so-called flow-of-information correlation} matrix \cite{audi1986, radon2000, litvinov2005, shubina2013}.

In order for the above method to work, a proper identification is required. The limitation of the TOF technique is the need for an extra tagging of the particles. Due to their non-destructive working principle, Schottky detectors allow for in-ring particle identification (PID) based on the simulation of peak position on the revolution frequency spectra \cite{freire2024thesis,freire2024zenodo}. This can be performed on the spectrum of a multi-component beam or by stacking of individual spectra in case of single ion injection. The latter is typically the case for the R3 storage ring at RIKEN, in fact this is a unique feature of the R3, which allows for the tagging and identification of individual particles.

\subsection{Advantages of SCD for R3 storage ring}

In the R3 storage ring, already very high mass measurement accuracies \yl{have been} achieved by the tagging method \cite{abe2024}. Here B$\rho$ determination and PID is carried out outside of the R3. While the additional path length adds to uncertainties, the PID process limits the transmission. Particle identification is generally done at \yl{focal plane (F3) of the BigRIPS separator} before the R3 storage ring, where energy loss is measured and plotted with repect the time of flight between sections F2 and F3. PPACs detectors are used for position measurements at the F5 section, but often they are used for particle identification if the resolution at section F3 is not adequate.

LISE++ simulations show that high-Z species such as \textsuperscript{216}Pb could gain up to a factor 100 in transmission efficiency, \yl{mainly due to charge exchange reactions,} if the PPACs are removed from the beam line, whereas mid-Z region particles such as \textsuperscript{78}Ge \yl{are} not much affected. At the same time, this very feature will come in handy during the benchmarking of the new SCD (accepted proposal NP2412-RIRING10) using mid-Z range particles: the PPAC detectors can stay in the beam line for an independent confirmation of the non-destructive in-ring PID. 

Finally, the performance of the SCD will show whether the overall gain in accuracy is comparable to existing methods. However, as RIKEN's world's highest intensities are currently affected by limited transmission, it is hoped that the SCD will make a significant contribution to RIKEN's nuclear physics programme.

\subsection{Future data acquisition systems}

The development of novel Schottky detectors goes hand in hand with the development of new data acquisition (DAQ) systems \cite{trageser2015}. Continuous and wide-band acquisition of signals from Schottky detectors places ever increasing demands on dedicated DAQs and processing hardware, which becomes a bottleneck to the scalability of such systems due to either their high initial price or vendor lock.

Software defined radio (SDR) based DAQ systems have proven to be an inexpensive and scalable alternative to such DAQ systems. The main aim of SDRs is to digitise measurable quantities as close as possible to their source (here: the Schottky detector) and perform all signal processing and analysis in the software domain using high level programming. Open hardware and open source SDRs allow for collaborative development and compatibility with Open Science standards such as F.A.I.R data principles of publicly funded research \cite{gofair,publiccode}.

SDR libraries such as GNURadio \cite{gnuradio} can be used with PID and other related libraries (see \cite{freire2024zenodo,sanjari2023}) for the purpose of monitoring and analysing Schottky spectra, with the aim of semi-automatic or automatic processing of incoming data.

\section{Summary and outlook}

Schottky cavity based detectors are fast and sensitive devices that can be used for mass and lifetime measurements of exotic isotopes. \yl{The attainable high mass resolving power allows for studying known or for searching new low lying isomeric states}. New position sensitive detectors, such as the newly installed SCD for the R3 storage ring at RIKEN, allow alternative interpretations by increasing the dimensionality of the acquired data. Open source data acquisition systems based on Software Defined Radio complement the operation of Schottky detectors and contribute to the scalability of data handling. Streams of time-stamped data samples from many Schottky detectors can be monitored and analysed together in order to extract features such as PID, mass and lifetime of hundreds of spectral lines as well as storage ring features such as $\alpha_p$ curve at the same time, giving birth to the idea of \textit{Software Defined \yl{Experiments for} Nuclear Astrophysics}. It is hoped that these developments will enable an exciting future for nuclear astrophysics research at both existing and future heavy ion storage ring facilities such as the ESR and Collector Ring (CR) at GSI/FAIR, R3 at RIKEN and SRing at the future HIAF facility.

\section{Acknowledgements}

We acknowledge financial support of MPIK Heidelberg. M.~S.~S., Yu.~A.~L. and J.~G. acknowledge support by the State of Hesse within the Research Cluster ELEMENTS (Project ID 500/10.006). E.~M. and Yu.~A.~L. acknowledge support by the project ``NRW-FAIR", a part of the programme ``Netzwerke 2021", an initiative of the Ministry of Culture and Science of the State of North Rhine-Westphalia.



\begin{thebibliography}{99}

\bibitem{schaaf1991} U. Schaaf, \textit{Schottky-Diagnose und BTF-Messungen an gekuehlten Strahlen im Schwerionen-Speicherring ESR}, PhD Thesis (Frankfurt: Goethe University Frankfurt 1991) (in German)

\bibitem{chattopadhyay1984} S. Chattopadhyay, CERN preprint, CERN-84-11 (1984)

\bibitem{caspers2008} F. Caspers, CERN preprint, CERN-2009-005: 407 (2008)

\bibitem{nolden2011} F. Nolden \textit{et al.}, Nucl. Instrum. Meth. A \textbf{659}(1):69–77 (2011)

\bibitem{sanjari2013} M. S. Sanjari \textit{et al.}, Phys. Scripta, \textbf{T156} (2013) 

\bibitem{wu2013}  J. X. Wu \textit{et al.}, Nucl. Instrum. Meth. B, \textbf{317}:623-628 (2013)

\bibitem{wang2025} Q. Wang \textit{et al.}, Nucl. Sci. Tech. \textbf{36}:17 (2025) 17

\bibitem{suzaki2015} F. Suzaki \textit{et al.}, Phys. Scripta. \textbf{T166}:014059 (2015) 

\bibitem{yamaguchi2016} Y. Yamaguchi \textit{et al.}, Proceedings of the International Workshop on Beam Cooling and Related Topics, FRYAUD01:182 (2015)

\bibitem{sanjari2020} M. S. Sanjari \textit{et al.}, Rev. Sci. Instrum. \textbf{91} (8):083303 (2020) 

\bibitem{litvinov2013} Yu. A. Litvinov \textit{et al.}, Nucl Inst Meth B, \textbf{317}:603–616 (2013) 

\bibitem{steck2020} M. Steck M., Yu. A. Litvinov, Prog. Part. Nucl. Phys. \textbf{115}:103811 (2020) 

\bibitem{hausmann2000} M. Hausmann \textit{et al.}, Nucl. Instrum. Meth. A, \textbf{446}:569--580 (2000)

\bibitem{hausmann2001} M. Hausmann \textit{et al.}, Hyperfine Interact. \textbf{132}:289--295 (2001)

\bibitem{wollnik2015} H. Wollnik, Hyperfine Interact. \textbf{235}:1--5 (2015)

\bibitem{zhang2016} Y. H. Zhang \textit{et al.}, Phys. Scripta \textbf{91}:073002 (2016)

\bibitem{yamaguchi2021} T. Yamaguchi \textit{et al.}, Prog. Part. Nucl. Phys. \textbf{120}:103882 (2021)

\bibitem{wang2023} M. Wang \textit{et al.}, Phys. Rev. Lett. \textbf{130}:192501 (2023)

\bibitem{litvinov2011} Yu. A. Litvinov, Rep. Prog. Phys. \textbf{74}:016301 (2011)

\bibitem{bosch2013} F. Bosch \textit{et al.}, Prog. Part. Nucl. Phys. \textbf{73}:84--140 (2013)

\bibitem{leckenby2024} G. Leckenby \textit{et al.}, Nature \textbf{635}:321–326 (2024)

\bibitem{sidhu2024} R. S. Sidhu \textit{et al.}, Phys Rev Lett \textbf{133}:232701 (2024)

\bibitem{kienle2013} P. Kienle \textit{et al.}, Phys. Lett. B \textbf{726}(4–5):638-645 (2013)

\bibitem{tu2018} X. L. Tu \textit{et al.}, Phys. Rev. C \textbf{97}:014321 (2018) 

\bibitem{freire2024} D. Freire-Fernández \textit{et al.}, Phys. Rev. Lett. \textbf{133}:022502 (2024)

\bibitem{sanjari2025} M. S. Sanjari \textit{et} al. Eur. Phys. J. A,  in prep. (2025)

\bibitem{freire2024thesis} D. Freire-Fernández, PhD Thesis, \textit{First nuclear two-photon decay measurements at storage rings}, (Heidelberg: University of Heidelberg 2024) (in English)

\bibitem{abe2024} Y. Abe \textit{et al.}, Nucl. Instrum. Meth. A \textbf{1072}:170083 (2025)

\bibitem{wang2025} Q. Wang, GSI GET Involved Program Report GSI Helmholtz Center Darmstadt (2025)

\bibitem{chen2014hist} X. Chen, GSI-Report 2014-3 (2014) 

\bibitem{sanjari2014} M. S. Sanjari et. al., Proceedings of the International Beam Instrumentation Conference, Monterey, CA, USA, WEPD03 (2014).

\bibitem{sanjari2015} M. S. Sanjari S. et. al., Phys. Scripta \textbf{T166}:014060 (2015).

\bibitem{dmytriiev2020} D. Dmytriiev \textit{et} al. Nucl. Instrum. Meth. B \textbf{463}:320-323 (2020)

\bibitem{wang2022} M. Wang \textit{et al.}, Phys. Rev. C \textbf{106}:L051301 (2022)

\bibitem{zhang2023} M. Zhang \textit{et al.}, Eur. Phys. J A \textbf{59}:27 (2023) 

\bibitem{fgm}
B. Franzke, H. Geissel, G. M{\"u}nzenberg, Mass. Spectr. Rev. \textbf{27}:428--469 (2008)

\bibitem{litvinov2023} Yu. A. Litvinov, Eur. Phys. J Web. Conf. \textbf{290}:02002 (2023)

\bibitem{hudsonchang2025} G. W. Hudson-Chang \textit{et al.}, these proceedings (2025)

\bibitem{chen2015} X. Chen\textit{et al.}, Hyperfine Interact. \textbf{235}:51–59 (2015) 

\bibitem{chen2016} X. Chen \textit{et al.}, Nucl. Instrum. Meth. A \textbf{826}:39-47 (2016)

\bibitem{audi1986} G. Audi\textit{et al.}, Nucl. Instrum. Meth. A \textbf{249}:443 (1986) 

\bibitem{radon2000} T. Radon \textit{et al.}, Nucl. Phys. A \textbf{677}:75–99 (2000)

\bibitem{litvinov2005} Yu. Al. Litvinov \textit{et al.}, Nucl. Phys. A \textbf{756}:3–38 (2005)

\bibitem{shubina2013} D. Shubina \textit{et al.}, Phys. Rev. C \textbf{88} 024310 (2013)

\bibitem{freire2024zenodo} D. Freire-Fernández, Zenodo \textbf{8172226} (2024)

\bibitem{trageser2015} C. Trageser C. \textit{et} al. J Phys. Conf. Ser. \textbf{635}:022085 (2015)

\bibitem{gofair} \url{https://www.go-fair.org/}, retrieved 7th April 2025

\bibitem{publiccode} \url{https://publiccode.eu/en/}, retrieved 7th April 2025

\bibitem{gnuradio} \url{https://github.com/fair-acc/gnuradio4}, retrieved 7th April 2025

\bibitem{sanjari2023} M S. Sanjari, Zenodo \textbf{7615693} (2023)

\bibitem{yu2024}
Y. Yu \textit{et al.}, Phys. Rev. Lett. \textbf{133}:222501 (2024)

\bibitem{geissel2005}
H. Geissel\textit{et al.}, J Phys. G \textbf{31}:1779-1783 (2005)

\bibitem{geissel2006}
H. Geissel \textit{et al.}, Hyperfine Interact. \textbf{173}:49--54 (2006) 

\end{thebibliography}
\end{document}